\documentclass[aps, prb, floatfix, notitlepage
]{revtex4-2}

\usepackage{amsmath,amssymb,mathrsfs}
\usepackage{psfrag}
\usepackage{graphicx}
\usepackage{graphics}
\usepackage{epsfig}
\usepackage{bm}
\usepackage{color}
\usepackage{verbatim,color,ulem}

\usepackage{amsmath}
\usepackage{here}
\usepackage{systeme}
\usepackage{xcolor}
\usepackage{multirow}
\usepackage{resizegather}
\usepackage{physunits}
\usepackage[utf8]{inputenc}
\usepackage[T1]{fontenc}
\usepackage{mathptmx}
\usepackage{etoolbox}
\usepackage{color}

\newcommand{\beq}{\begin{equation}}
\newcommand{\eeq}{\end{equation}}
\newcommand{\beqa}{\begin{eqnarray}}
\newcommand{\eeqa}{\end{eqnarray}}

\makeatletter
\def\@email#1#2{%
 \endgroup
 \patchcmd{\titleblock@produce}
  {\frontmatter@RRAPformat}
  {\frontmatter@RRAPformat{\produce@RRAP{*#1\href{mailto:#2}{#2}}}\frontmatter@RRAPformat}
  {}{}
}
\makeatother
\begin{document}

\title{Quantum Vortices in Curved Geometries}

\author{A. Tononi$^1$,  L. Salasnich$^{2,3}$,
and A. Yakimenko$^{2,4}$}

\address{$^{1}$ ICFO-Institut de Ciencies Fotoniques, The Barcelona Institute of Science and Technology, Av. Carl Friedrich Gauss 3, 08860 Castelldefels (Barcelona), Spain \\
$^{2}$Dipartimento di Fisica e Astronomia ``Galileo Galilei'',
Universit\`a di Padova, and INFN, Sezione di Padova,
Via Marzolo 8, 35131 Padova, Italy \\
$^{3}$Istituto Nazionale di Ottica (INO) del Consiglio Nazionale
delle Ricerche (CNR), Via Nello Carrara 1, 50019 Sesto Fiorentino, Italy \\
$^{4}$Department of Physics, Taras Shevchenko National University
of Kyiv, 64/13, Volodymyrska Street, Kyiv 01601, Ukraine}

\begin{abstract}
The control over the geometry and topology of quantum systems is crucial for advancing novel quantum technologies. This work provides a synthesis of recent insights into the behaviour of quantum vortices within atomic Bose-Einstein condensates (BECs) subject to curved geometric constraints.
 We highlight the significant impact of the curvature on the condensate density and phase distribution, particularly in quasi-one-dimensional waveguides for different angular momentum states.
An engineered periodic transport of the quantized vorticity between density-coupled ring-shaped condensates is discussed. The significant role of curved geometry in shaping the dynamics of rotational Josephson vortices in long atomic Josephson junctions is illustrated for the system of vertically stacked toroidal condensates. Different methods for the controlled creation of rotational Josephson vortices in coupled ring systems are described in the context of the formation of long-lived vortex configurations in shell-shaped BECs with cylindrical geometry. Future directions of explorations of vortices in curved geometries with implications for quantum information processing and sensing technologies are discussed.

\end{abstract}

\maketitle

\section{Introduction}
In the last few years, the superfluidity of atomic Bose-Einstein condensates (BECs) confined in low-dimensional curved geometries attracted increasing interest in the ultracold atoms community. Explorations of this kind aim to identify phenomena which are absent (or just occur differently) in low-dimensional flat systems. The starting point for reviewing such investigations is identifying the features which characterize a geometry as \textit{curved}. These include: the local curvature, the periodic boundaries, the manifold compactness, and the non-trivial topology \cite{2023NatRP...5..398T}. Their combination generates examples of one- and two-dimensional geometries such as rings, curved wave guides, spheres, tori, etc.: analyzing atomic gases confined in these manifolds can clarify how superfluidity and vortex physics interplay with the hosting geometry
\cite{2012PhRvE..85c1150K,2016RSPSA.47260271S,2017JETPL.105..458R,2018PhRvE..97a2117G,2019JPhA...52U5501B,2019PhRvL.123p0403T,2020JMMM..51367254M,2022PRXQ....3a0316L,2022PhRvA.105b3324T,2022PhRvA.105f3325D,2022PhRvR...4a3122T,2023PhRvA.108e3303H,2023PhRvA.108e3323Y,2024NJPh...26a3035B,2024PhRvA.109a3301W,2024PhysRevResearch.6.013133}.
Further explorations along this line will, in perspective, enable the possibility of engineering and controling superfluidity by geometric means, with possible technological and fundamental consequences.

The topological nature of quantum vortices is essential for understanding their stability and dynamics in superfluid BECs. In a BEC state, bosonic atoms condense into the same single-particle state, exhibiting macroscopic quantum phenomena. A quantum vortex in a BEC is a region where the phase of the wave function winds around a line or point, with the density of the condensate going to zero at the core of the vortex. The circulation of the velocity field around the vortex is quantized, namely, it takes discrete values corresponding to the vortex winding number or topological charge. Experiments with BECs that create and manipulate quantum vortices provide empirical evidence of their topological nature \cite{Chevy2008Vortices}. In particular, \textcolor{black}{this property affects}
the stability of the quantum vortices for substantial perturbations and their dynamics during collisions (e.g., vortex-antivortex annihilation).
Thus, quantum vortices in a curved geometry are a unique platform to study the intimate relation between topology and curvature.

From the mathematical viewpoint, a deep connection between the curvature and the topological properties of a surface is established by the Gauss-Bonnet theorem \cite{1993rgmi.book.....C}. For a given compact and boundaryless manifold, like the surface of a sphere or of a torus, the theorem states that the surface integral of the Gaussian curvature is equal to $2\pi \chi$. Here, the Euler characteristic of the surface $\chi$ is a topologically-invariant integer number which provides information on the manifold holes, connected components, and other topological properties.
The theorem provides therefore a connection between apparently distinct areas of mathematics such as differential geometry and topology.
When analyzing the physics of superfluids, this mathematical result implies that a different vortex dynamics in flat versus curved superfluids cannot always be attributed either to curvature or to topology (or also to the boundary conditions in the case of open manifolds). In these cases, one should then refer to the general role of the curved geometry, in opposition to a flat surface geometry with trivial topology. The general message is that the properties of the surface combined with the topological nature of the phase field influence the presence and the configuration of vortices. Understanding better this interplay in specific cases is crucial for advancing in predicting and in explaining the behavior of vortices in relation to the geometric and topological properties of the underlying space.

Vortices in curved geometries have been extensively investigated across various physical systems, becoming the focal point of numerous theoretical and experimental studies. Their analysis in ultracold atoms systems has developed over the years and we highlight in this paper a few recent results connected to vortex physics.
In particular, the intricate behavior of quantum vortices in curved geometries \cite{2017PhRvA..96f3608G, 2019PhRvA..99f3602M, 2020PhRvA.101e3606G, PhysRevA.105.023307,2021PhRvA.103e3306B} \textcolor{black}{points towards} novel quantum states and phase transitions, governed by the interplay of curvature, interaction, and topological constraints. The authors of Ref.  \cite{2017PhRvA..96f3608G} have studied superfluid vortex dynamics on an infinite cylinder, highlighting the unique behaviour due to quantum-mechanical effects, such as quantized translational velocities for vortices due to the single-valued requirement of the condensate wave function, and extends the analysis to finite cylinders.
The authors of Ref. \cite{2021PhRvA.104f3310R} have studied the thermodynamics of shell-shaped BECs in bubble traps, showing how tuning trap parameters shifts the system from solid spheres to thin shells, affecting the excitation spectrum, and critical temperature for condensation.
The study \cite{2020PhRvA.102d3305P} investigates vortex dynamics in shell-shaped BECs, particularly focusing on the stabilization of vortex-antivortex pairs under rotation, and highlights how these dynamics can serve as a nondestructive method to characterize and distinguish hollow BECs from filled ones.

\textcolor{black}{The experimental observation of these phenomena requires to confine superfluids in curved geometries.
For this scope, the work} \cite{2019npjMG...5...30L} proposed a framework for creating BECs in shell geometries by using radiofrequency dressing, aiming to explore new collective modes and topological effects in microgravity conditions facilitated by the NASA Cold Atom Laboratory on the International Space Station. Investigations conducted aboard the International Space Station \cite{2022Natur.606..281C}
have advanced our understanding of ultracold atomic systems through the observation of bosonic gases confined in ellipsoidal bubbles. These experiments, using the unique conditions of microgravity, have explored novel geometric and topological configurations of quantum systems. Recent experimental efforts \cite{PhysRevLett.129.243402}
have successfully realized shell-shaped BECs on Earth, utilizing dual-species mixtures composed of sodium and rubidium atoms. This was achieved within a specifically engineered optical dipole trap, facilitating the formation of a \textcolor{black}{core-shell structure.
These two realizations currently feature competing advantages towards the goal of investigating superfluid vortex dynamics in \textit{two-dimensional} (thin) \textit{condensate} shells: while the microgravity realization reached the \textit{two-dimensional} closed-shell confinement of thermal gases, the dual-species mixture could instead produce three-dimensional (thick) shells but with \textit{no detectable thermal density}.
In perspective, the mixture realization could also be suited for analyzing the dynamics in curved geometries of massive-core vortices \cite{2020PhRvA.101a3630R} which, in this context, are topological excitations of the superfluid constituting the shell whose cores are filled by the atoms of the other species.
In addition, a} novel approach \cite{2022NJPh...24i3040G} to preparing a quantum gas in a shell trap has been demonstrated, starting from a degenerate Bose gas in a hybrid trap that combines a magnetic quadrupole trap with an attractive optical trap. Experimentally accessible methods \cite{2022JAP...132u4401R} allow efficient transitions of quantum gas into a shell-shaped trap with minimal excitation of the centre-of-mass motion.
\textcolor{black}{
Finally, relevant to the theoretical analyses of the present work which focus on ring-like geometries, we point out the implementations of atomic confinement in rings \cite{PhysRevA.74.023617,Heathcote_2008,2021JPhB...54l5302D}, obtained by intersecting a magnetic shell potential with optical traps that confine the gas along the shell equator.
}

\textcolor{black}{
Recent advancements in experimental physics have led to significant breakthroughs in the understanding and manipulation of BECs, ranging from the exploration of novel quantum states in microgravity to the development of sophisticated atomtronic circuits \cite{2021AVSQS...3c9201A} on Earth.
 The development of a painting potential technique \cite{2009NJPh...11d3030H, 2015NJPh...17i2002R}  employing rapidly moving laser beams to create dynamic, arbitrary, and stable optical dipole potentials enables the generation of BECs in various geometrical configurations, including toroids and ring lattices. This approach offers the capability to dynamically alter the trapping potential and allows for the generation, guidance, and manipulation of coherent matter waves within a single device, akin to an integrated optical circuit. Thus, it opens prospects for the realization of sophisticated quantum computing and sensing devices. Using strongly correlated neutral bosons in a mesoscopic ring-shaped potential, the work \cite{SciPostPhysPerrin} uncovers the fractionalization of angular momentum, a quantum-many-body feature, enabling the operation of rotation sensors and gyroscopes with extremely high sensitivity. Recent investigation \cite{10.21468/SciPostPhys.15.4.128} demonstrates the use of a toroidal BEC as a rotation sensor by measuring the interference between counter-propagating azimuthal phonon modes.
}

Curved geometries play a crucial role across various physical systems, extending well beyond atomic BECs.
Several studies discussed the role of curvature and diverse geometric aspects in condensed matter systems in the last decades \cite{2010RvMP...82.1301T, 2002dgcm.book.....N, Streubel_2016,Peng2020StrainEngineering, Gentile2022, Grass2024}.
One notable example is in the superconductors, specifically in superconducting nanoshells. The paper \cite{PhysRevB.79.134516} demonstrate the coexistence of the Meissner and vortex states on a nanoscale superconducting spherical shell demonstrating the significant impact of curved geometries. In these systems, the curvature and topology of the superconducting layer influence the behavior of vortices and antivortices, leading to phenomena such as the separation of vortex-antivortex pairs and the formation of a Meissner belt around the equator of the sphere.
Non-planar geometries give rise to exotic quantum states, yet their realization within solid materials poses significant challenges. An interesting approach was demonstrated in Ref. \cite{2022PRXQ....3a0316L}, which employs an atomic quantum simulator allowing atoms to mimic electronic behaviour in materials, within a curved, cylindrical space subjected to a radial magnetic field.
Furthermore, curved geometries are essential in other areas of physics, such as general relativity, where the curvature of spacetime governs gravitational interactions, and in soft matter physics, where the curvature of interfaces can influence the behaviour of colloids, droplets, and biological membranes. These examples underline the broad relevance of curved geometries in physics, extending far beyond atomic BECs.

Providing a comprehensive literature overview on this topic exceeds the scope of this short review-perspective paper; however, a thorough analysis of the literature is available in several review papers.
In particular, review \cite{2010RvMP...82.1301T} presents a comprehensive overview of how topological defects in the form of vortices and phase dislocations interact with curved geometries, particularly in thin film systems.
The work \cite{2002ApMRv..55B..15N} extensively covers the Hamiltonian dynamics of vortex motion in classical fluid mechanics.
Recent review paper \cite{2023NatRP...5..398T} presents a comprehensive overview of shell-shaped bosonic gases, as well as the behaviour of both thin and thick spherical shells of bosonic gases.
The paper \cite{2023NatRP...5..398T} also highlights the unique properties of low-dimensional quantum gases in curved geometries.
Another recent review paper \cite{TONONI2024} extensively explores the quantum statistical physics of two-dimensional shell-shaped quantum gases at zero and finite temperatures.
It overviews BECs, superfluidity, vortex physics, and the crossover from BCS to BEC regimes, particularly emphasizing the unique aspects brought by the curved geometry. The paper also discusses hydrodynamic excitations and their connection to the Berezinskii-Kosterlitz-Thouless (BKT) transition, positioning shell-shaped atomic gases as a promising experimental platform for studying quantum many-body physics in curved geometries.

In this work, we provide a concise overview of the recent findings on vortices in atomic BECs related to distinct curved geometries: (i) curved ring-shaped waveguides, which can be effectively modelled as 1D or 2D systems (Section II), and (ii) coupled curved waveguides exhibiting essentially 3D dynamics of the vortices influenced by curved geometries (Section III).
In Subsection II A, we consider particular examples of the essential impact of curved geometry on the properties of quantum vortices. Specifically, we examine how curvature affects the density and phase distribution of the condensate in a quasi-1D waveguide across various angular momentum states. In Subsection II B
we discuss the role of curvature in deterministic transition between quantum states with different angular momentum.  We consider an example of an experimentally accessible method for the controllable creation of quantum vortices in quasi-2D single ring-shaped condensates.
In Subsection II C we introduce physical systems that hold promise as platforms for novel quantum sensors: a dual-ring setup consisting of interconnected rings.
Subsection III A highlights the significance of curvature in the 3D dynamics of rotational Josephson vortices in long atomic Josephson junctions formed by a 3D system of vertically stacked toroidal condensates.
In Subsection III B we focus on the methods of controllable formation of Josephson vortices in a system of stacked rings. Subsection III C discusses remarkable 3D hybrid vortex configurations illustrating properties of the quantum vortices in a cylindrical trap.
Section IV includes a summary and analysis of the prospects of future studies related to vortices in curved geometries.


\section{Quantum vortices in curved wave guides}
\textcolor{black}{Due to the intrinsic curvature of atomtronic circuits' wave guides \cite{1977JChPh..67.3061S,PhysRevLett.124.250403,Campo2014}, mastering the manipulation of condensate density and phase distribution becomes essential for advancing quantum sensors and information processing devices utilizing atomic BECs.} Enhanced control over these aspects in curved waveguides opens a way for creating novel quantum devices, offering improved sensitivity and multifunctionality. In the following section, we consider several examples of remarkable dynamical properties of the quantum vortices driven by curvature geometry in quasi-1D and 2D curved atomic waveguides. In the toroidal topology of the trapped BEC under consideration, vortices are inherently pinned at the center of the ring-shaped condensate. This pinning occurs where the vortex energy reaches a local minimum, thereby ensuring that even multi-charged ($q > 1$) metastable vortex states exhibit significant robustness. This stability allows considering strong variations of the curved geometry and investigating induced phase and density transformations in the quasi-1D and quasi-2D toroidal BECs.


\subsection{Curvate-induced control on density and phase in curved waveguides.}\label{subsec:eliipse}
\begin{figure}[htb]
\centering
\includegraphics[width=\textwidth   ]{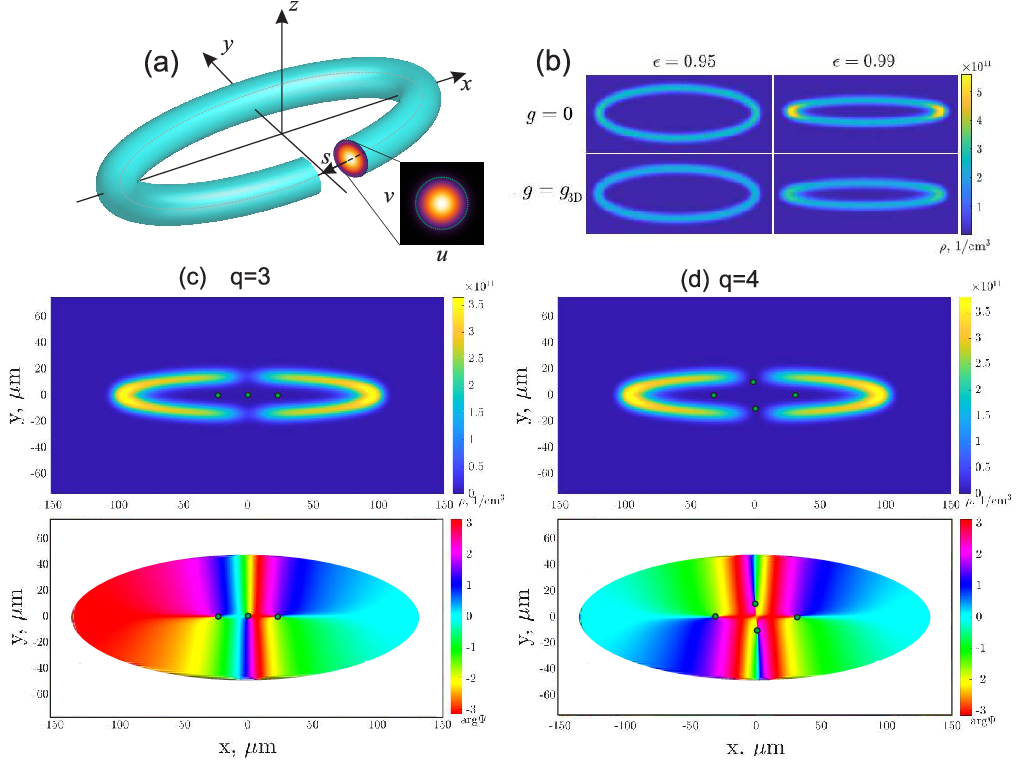}
\caption{(a) Schematics of the elliptic waveguide geometry used for trapping a condensate as described in Ref. \cite{2023NJPh...25j3003N}. Shown the 3D plot of the density isosurface and
condensate density in a perpendicular cross-section (brighter colours indicate higher condensate density). (b) Density
distribution at the $z=0$ plane for the non-interacting case (upper row) and the case with repulsive interaction (lower row)\textcolor{black}{, shown for two different values of the eccentricity $\epsilon$}.
Notably, density modulations arise due to curvature in regions with high curvature, and these modulations are reduced by strong
repulsive interactions. (c) Colour-coded phase of the condensate in $z = 0$ plane combined with density isolines for the stationary states with superflow winding number  $q = 3$. (d) Winding number  $q = 4$. }
\label{fig:Curved_Ellipse}
\end{figure}

In our recent investigation \cite{2023NJPh...25j3003N}, we explored the influence of ellipticity-induced curvature on atomic BECs within quasi-1D closed-loop waveguides. Our theoretical framework reveals a remarkable interplay between curvature and inter-particle forces. It turns out that curvature-induced density modulations are significantly diminished in the presence of strong repulsive interactions. In areas of minimal curvature, particularly in waveguides exhibiting superflow with large eccentricity, we detect notable phase accumulation. \textcolor{black}{Furthermore}, in waveguides hosting vortices, we observed dynamic transitions across states characterized by distinct angular momenta.  The observed intricate relationship between curvature and interaction effects allows the engineering of quantum states within constrained settings with curved geometry.

The Gross-Pitaevskii equation (GPE) provides an accurate description of the dynamics of BECs in mean-field approximation. In three dimensions, the GPE is formulated as:
\begin{equation}
i\hbar\frac{\partial \Psi }{\partial t} = \left(-\frac{\hbar^2}{2m} \nabla ^{2}+V_{\text{ext}}(\mathbf{r},t)+g|\Psi |^{2} \right) \Psi, \label{GPE3Ddim}
\end{equation}
where, $\Psi(\mathbf{r},t)$ is the wave function of the condensate.
In Ref. \cite{2023NJPh...25j3003N} we have considered two scenarios: the non-interacting condensate with $g=0$, and the repulsive interaction case with $g=g_{3D}=4\pi a_{s}\hbar^{2}/m$, where $a_s$ is the $s$-wave scattering length.
The wave function satisfies the normalization condition $\int|\Psi|^2 d\mathbf{r}=N$, where $N$ is the number of particles in the condensate.


We model the external trapping potential $V_\mathrm{ext}(x,y,z)$ as a combination of an parabolic potential in the $z$-direction and an elliptic waveguide in the $(x,y)$ plane, with a larger semi-axis $a$: 
\begin{equation}\label{Eq:V_ext}
V_\mathrm{ext}(x,y,z)=\frac12 m\omega_z^2 z^2+\frac12 m\omega_\perp^2 R^2(x,y).
\end{equation}
Here, $R(x,y)=\left[(x-x_0)^2+(y-y_0)^2\right]^{1/2}$ characterizes the minimum distance between the point in $(x,y)$ plane and a coplanar point $(x_0,y_0)$ at the ellipse.

In order to isolate and examine the influence of curvature on the condensate density distribution, maintaining a uniform cross-section along the waveguide is crucial. This requirement is fulfilled by the employed trapping potential described by Eq. (\ref{Eq:V_ext}), which establishes a parabolic trap $z$-direction and a waveguide in $(x,y)$ plane with a parabolic profile perpendicular to the ellipse direction. The isolines of the potential form circles of constant radius along the quasi-1D waveguide, as shown in Figure \ref{fig:Curved_Ellipse} (a).

We have found a steady-state solution of the form: $    \Psi(\textbf{r},t)=\Tilde{\Psi}(\textbf{r})e^{-i\mu t/\hbar},$
where $\mu$ is the chemical potential. In general, the complex wave function $\Tilde{\Psi}=|\Tilde{\Psi}(\textbf{r})|e^{i\Phi(\textbf{r})}$ exhibits an inhomogeneous phase $\Phi(\textbf{r})$ with a circulation given by
\begin{equation}
    \label{eq:Circulation}
    \oint_C \nabla\Phi(\textbf{r})\cdot d\textbf{l} =2 \pi q,
\end{equation}
where the contour $C$ represents the ellipse $x_0^2/a^2+y_0^2/b^2=1$, and $q$ is an integer denoting the topological charge of the wave function. For the ground state, $q=0$, while $q>0$ corresponds to a state with $q$ vortices, resulting in a counter-clockwise flow in the waveguide.

Figure \ref{fig:Curved_Ellipse} (b) illustrates examples of the density distributions for the ground states ($q=0$) in the non-interacting case ($g=0$) and the repulsive interaction case ($g=g_{3D}>0$).
To compare the properties of the numerically obtained ground states, determined by solving the 3D stationary GPE, with an approximate \textcolor{black}{ effective potential} induced by curvature, we have used in Ref. \cite{2023NJPh...25j3003N} a non-polynomial  Schr\"odinger equation (NPSE)  \cite{SciPostPhysCore}. This model incorporates a quantum curvature-induced potential, which exhibits a double-well shape. The eccentricity of the ellipse notably influences the specific features of this potential.

The use of an elliptic waveguide enables the accumulation of a
substantial phase jump within a localized region, a remarkable achievement inaccessible
in a single-connected quasi-one-dimensional condensate.  Figure \ref{fig:Curved_Ellipse} presents examples of density and phase distributions for stationary states with $q = 3$ in \ref{fig:Curved_Ellipse} (c) and $q = 4$ in \ref{fig:Curved_Ellipse} (d) for an eccentricity of $\varepsilon=0.99$. According to findings in Ref. \cite{2023NJPh...25j3003N}, the spatial distribution of the phase transitions from a uniform phase gradient in a ring of constant curvature \textcolor{black}{(so that the eccentricity, $\epsilon=0$)} to an essentially non-uniform phase distribution in waveguides with variable curvature. This property suggests that an increase in eccentricity leads to a more homogeneous phase distribution in areas of greater curvature, thus localizing the phase variation in areas of small curvature.

The curvature effect in the elliptic waveguide significantly enhances the localization of a substantial phase shift.
This is particularly evident when $q\ge 3$ (as illustrated in Fig. \ref{fig:Curved_Ellipse}), where the pronounced phase variation leads to an essential redistribution of density. Such a phase discontinuity can lead to the emergence of regions with density nodes, effectively acting as domain walls or dark solitons that separate areas with a phase difference of $\pi$.

Comprehensive numerical simulations based on the damped Gross-Pitaevskii equation have revealed complex behaviours in the system as detailed in Ref. \cite{2023NJPh...25j3003N}. It turns out that not only the ground state ($q = 0$) but also the single-charged ($q = 1$) and double-charged ($q = 2$) superflows remain stable over long-term evolution, even at high eccentricity ($\epsilon = 0.99$), maintaining their coherent flow patterns without significant transformations.
For higher-charged superflows with $q \ge 3$, a series of complex dynamic transformations among different states were observed. These superflows undergo significant changes in their flow patterns and topological structures. The evolution of these superflows demonstrates transitions between various states, leading to a complex and intricate dynamic landscape, as was found in Ref. \cite{2023NJPh...25j3003N}.


\subsection{Dyanamics of persistent current formation in quasi-2D curved wave guides}\label{subsec:stiring}
\begin{figure}[htb]
\centering
\includegraphics[width=\textwidth   ]{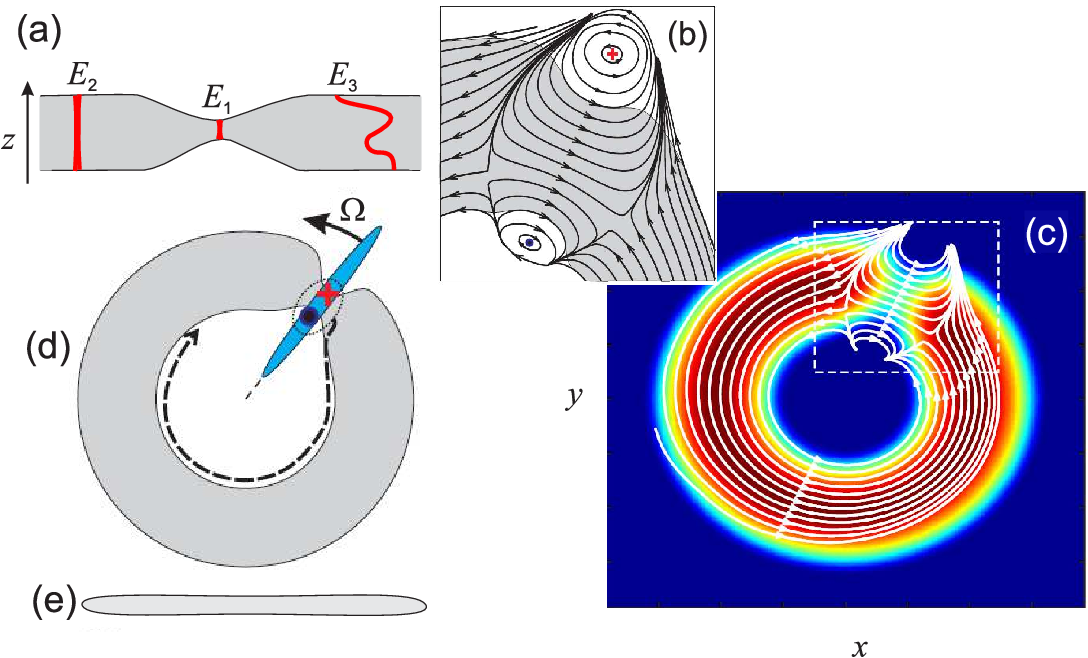}
\caption{Formation of a persistent current in a quasi-2D toroidal BEC by a rotating weak link as described in Ref. \cite{PhysRevA.91.033607}. (a) Side view of the condensate near the weak link with embedded vortex lines (red curves depict vortex cores of three vortices). The quasi-2D geometry makes the vertical orientation of the vortex line energetically preferable so that the bent vortex line has higher energy than the vertical vortex line: $E_3>E_2$. The vertical vortex within the weak link resides in a lower-density area, resulting in even lower energy ($E_1<E_2<E_3$). (b) The top view of the superflow streamlines around the weak link, identifying a vortex and an antivortex with a red cross and a blue circle, respectively. (c) Top view with colour-coded density combined with the superflow streamlines obtained in Ref. \cite{PhysRevA.91.033607}. Schematics of the (d) top view and (e) side view of the pancake-shaped toroidal condensate. The phase slip is driven by a vortex line from the outer edge and an antivortex line from the inner region, which form a moving vortex-antivortex pair.}
\label{fig:Stirring}
\end{figure}

\textcolor{black}{Significant experimental advancements in creating atomic gases within toroidal geometries has opened novel prospects for the
studies of the fundamental properties of quantum vortices. The study of ring-shaped BECs has thus become a prominent focus in both experimental and theoretical research \cite{PhysRevLett.99.260401,PhysRevLett.111.205301,2020NatCo..11.3338R,PhysRevA.86.013629,PhysRevA.80.021601,PhysRevA.81.033613,PhysRevLett.106.130401,2001JPhB...34L.113B,PhysRevA.74.061601,PhysRevLett.110.025301,PhysRevA.88.051602}.}
The series of the experiments  \cite{PhysRevLett.110.025302,PhysRevA.88.063633,2014Natur.506..200E} has demonstrated deterministic, abrupt transitions between quantized circulation states in a ring-shaped Bose-Einstein condensate stirred by a rotating barrier. The stirring repulsive potential produced a moving weak link, i.e. a curved region on the ring with lower atomic density, as illustrated in Fig. \ref{fig:Stirring}. These quantum transitions are very sensitive to the amplitude and angular velocity of the weak link which might enable novel super-sensitive rotation sensors that could dramatically improve the accuracy of inertial navigation systems.  The experimental discoveries inspired further theoretical investigations which have addressed the role of curvature of the trapping geometry on phase slip dynamics.

In this subsection, we briefly overview our previous theoretical studies \cite{PhysRevA.91.033607,PhysRevA.91.023607}, inspired by the experimental findings reported in \cite{PhysRevLett.110.025302,PhysRevA.88.063633}. These phase slips, initiated by vortex excitations due to a rotating weak link, are crucial for understanding the interactions between localized condensate density reductions, atomic superflows, and the energetic and dynamic stability of vortices in the annular condensate.

The emergence of persistent currents is driven by a wide rotating barrier (refer to Fig. \ref{fig:Stirring}). This rotation not only modifies the density of the condensate but also establishes a weak link and excites superflows within the annular condensate.  \textcolor{black}{It is important, that repulsive barrier substantially modifies a geometry of the isolines of constant condensate density and simultaneously modify the local behaviour of the superflow, as is seen from side and top views of the weak link region. Indeed, in the weak link region the flow's streamlines and condensate surface are curved which substantially affect the energetic stability and dynamic behaviour of vortices in a stirred toroidal condensate  \cite{PhysRevA.91.033607}.}

To analyze the temporal evolution of the condensate we used the damped 3D GPE:
\begin{equation}\label{GPE}
(i-\gamma) \hbar \frac{\partial \tilde\Psi(\textbf{r},t)}{\partial t} = \left[\hat H + \tilde g |\tilde\Psi(\textbf{r},t)|^2 -\mu\right]\tilde\Psi(\textbf{r},t),
\end{equation}
where $\gamma$, is a small phenomenological dissipation parameter $\gamma\ll 1$. The Hamiltonian, denoted by $\hat H=-\frac{\hbar^2}{2M} \Delta + V(\textbf{r},t)$, incorporates the Laplace operator $\Delta$. The coupling strength is represented by $\tilde g = 4 \pi \hbar^2 a_s/M$, where $M$ is the mass of a $^{23}$Na.
We assume the condensate's temperature remains significantly below the condensation threshold, i.e., $T\ll T_c$. The chemical potential $\mu(t)$ in our dynamic simulations is dynamically adjusted to simulate an observed in experiments decay in the number of condensed particles over time, following $N(t)=N(0)e^{-t/t_0}$, with 
$t_0$ indicating the BEC's $1/e$ lifetime.

The external potential $V(\textrm{\textbf{r}},t)=V_t(r,z)+V_b(\textrm{\textbf{r}},t)$ involves the axially-symmetric, time-independent toroidal trap and the time-dependent potential of a rotating repulsive barrier $V_b(\textrm{\textbf{r}},t)$.
The trap is approximated by a harmonic potential
\begin{equation}\label{toroidal_trap}
V_{\textrm{t}}(r,z)=\frac12 M\omega_r^2(r-R)^2+\frac12 M\omega_z^2 z^2.
\end{equation}
with  $r=\sqrt{x^2+y^2}$.
 We assume a tunable weak link described by a barrier potential $V_b(\textbf{r}_\perp,t)$, which is homogeneous in a radial direction across the toroidal condensate,
\begin{equation}\label{straigthBeam}
V_b(\textbf{r}_\perp,t)=U(t)\Theta(\textbf{r}_\perp\cdot \textbf{n})e^{-\frac{1}{2c^2}\left[\textbf{r}_\perp\times \textbf{n}\right]^2},
\end{equation}
where  $\textbf{r}_\perp=\left\{x,y\right\}$ is the radius-vector in $(x,y)$ plane; the unit vector $\textbf{n}(t)=\left\{\cos(\Omega t),\sin(\Omega t)\right\}$  points along the azimuth of the barrier maximum. All parameters of the potential, including the time-dependent barrier amplitude $U(t)$, were chosen to match the experimental conditions in \cite{PhysRevLett.110.025302} (see Ref. \cite{PhysRevA.91.033607} for details). In the experiment \cite{PhysRevLett.110.025302} the values of angular velocity were taken
in the range that corresponds to linear velocities well below the speed
of sound propagating around the ring. For slow rotation rates pronounced quantized phase slips were observed at well-defined critical angular velocities.

Let us discuss the role of curved geometry in the mechanism of these phase slips. During the stirring procedure
two vortices of opposite charges are nucleated on the inner and outer edges of the weak link (see Fig. \ref{fig:Stirring}) with significant curvature. As the superflow velocity is bigger on the outer edge, the
vortex from outside enters the weak link first and moves towards the central hole as illustrated
in Fig. \ref{fig:Stirring} (b). The schematics Fig. \ref{fig:Stirring} (a) of the density distribution in the vertical direction in the vicinity of the weak link illustrates that the vortex within the weak link resides in a lower-density area, resulting in reduced energy.  While the vortex traverses the weak link, a negatively-charged
anti-vortex line from the central hole and the incoming vortex approach each other
and create a vortex-antivortex dipole. This coupled vortex-antivortex pair circles clockwise
inside the central hole until it reaches the region of the weak link. This moving vortex dipole
usually escapes from the central hole and finally decays.
If the barrier rotation rate is well above a threshold, usually a series \textcolor{black}{of} vortices
enter into the central hole and form bound pairs with anti-vortices from the central
hole. Finally, the vortex-antivortex pairs jump out of the condensate. Note that the total angular momentum of the escaping vortex dipole is equal to zero, but each time an
external vortex enters the condensate it adds one unit of topological charge to the persistent current in the ring.

In the experiment  \cite{PhysRevLett.110.025302} with oblate toroidal condensate dynamics of the vortices does not involve bending of the vortex line since the vertical position of the vortex is energetically preferable, as illustrated in Fig. \ref{fig:Stirring} (a). Thus the vortex dynamics can be treated as quasi-2D, allowing considering only motions in the horizontal plane, while the rest of the degrees of freedom remain frozen in the state with the lowest energy. The rotating weak link creates additional curvature in the condensate distribution in the horizontal plane, which drives the formation of the vortex dipole, as shown in Fig. \ref{fig:Stirring}. Furthermore, the weak link deforms the density distribution in the vertical direction which drives the vortex line to the weak link region as this position is energetically preferable, as is seen from Fig. \ref{fig:Stirring} (a).

The theoretical predictions of Ref. \cite{PhysRevA.91.033607} are in qualitative agreement with the experimental findings:
the threshold angular velocity decreases when the barrier height increases.
However, a comparison of the experimental series for the fixed barrier height shows that the theoretically predicted
 is higher than the experimentally
measured angular velocity for the phase slip $0\to 1$. It is worth mentioning, that accounting for the stochastic thermal effects \cite{PhysRevA.94.063642} and decreasing number of atoms \cite{Yakimenko2015} are not able to eliminate the discrepancy between theory and experimental observations. Thus, further theoretical and experimental work
is needed to describe quantitatively the phase slips observed in the experiments.


\subsection{Transfer of quantum vortices in co-planar linked atomic rings}\label{subsec:quantum_trigger}

\begin{figure}[htb]
\centering
\includegraphics[width=\textwidth   ]{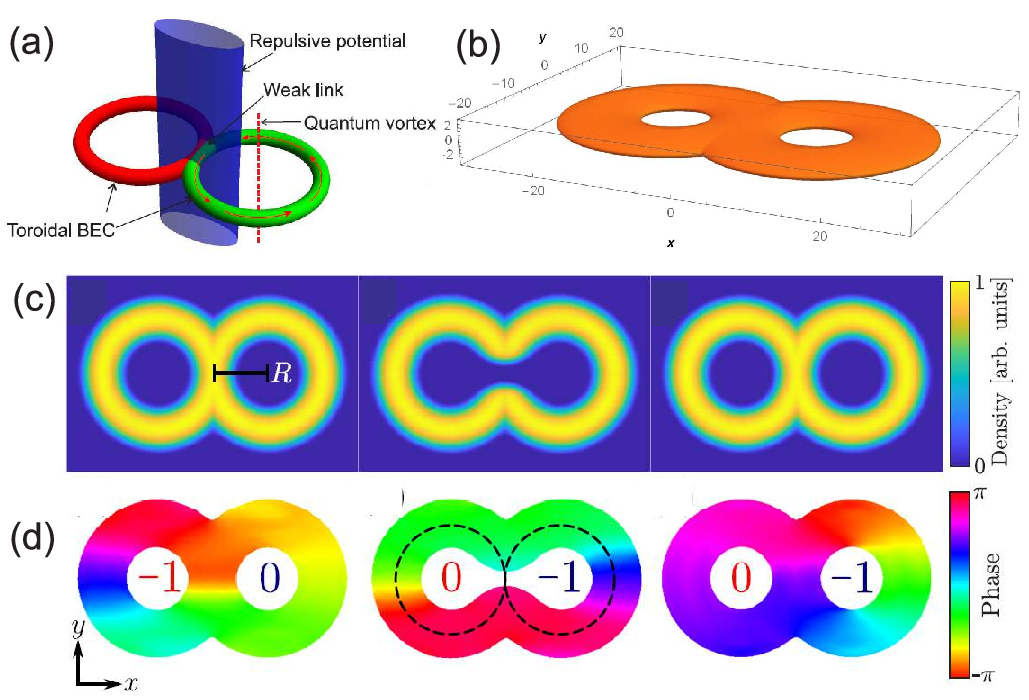}
\caption{(a) Schematics of quantum vortex trigger with the vortex line trapped in one of the rings. The quantum vortex
transfers through weak links between rings. (b) Condensate density isosurface for the double-ring system.  Persistent current oscillations showing (c) density and (d) phase. The phase shows an initial antivortex (clockwise circulation) imprinted into the left ring. After opening the gate between two
rings, the antivortex can periodically transfer as described in Ref. \cite{PhysRevResearch.4.043171}.}
\label{fig:Double_ring_planar}
\end{figure}
The recent work \cite{PhysRevResearch.4.043171} explores the dynamics of a BEC within a system of two co-planar ring-shaped structures, linked by a tunable, weak connection. This connection facilitates the coherent transfer of quantum vortices between the two rings. Recent investigations \cite{PhysRevResearch.4.L022038} of similar systems at the microscopic level revealed quantum phase slips in a double-ring framework featuring an adjustable weak link. While mean-field theory offers a view of a many-particle system that is both highly controllable and is not highly sensitive to thermal and quantum perturbations, the many-body perspective introduces the possibility of superposition and entanglement among these current states. Within this framework, an additional ring could serve either as a non-invasive diagnostic tool for experiments in the primary ring or as part of a novel, dual-ring setup. In the latter case, the system's sensitivity to external variables during the transfer of the persistent current could be applied for high-precise measurements.

A state with asymmetric distribution of the angular momentum in a system of linked rings can be created in a similar way, as it was realized experimentally: by phase imprinting, by stirring, and by stochastic merging. An interesting alternative was suggested in Ref.  \cite{Bland_2020}, which demonstrated the spontaneous emergence of persistent currents and vortices in a system of two co-planar ultracold bosonic gas rings with a common interface or in a lemniscate geometry, by quenching across the Bose-Einstein condensation phase transition, revealing the potential for independent winding numbers and flow directions in the rings, with their persistence supported by domain formation's local nature and the topological protection of winding numbers.

The consistent transfer of persistent current between two rings was theoretically investigated in Ref. \cite{PhysRevResearch.4.043171}. The comprehensive simulations, spanning both 2D and 3D within a quasi-2D framework, have shown that these oscillations persist over time, even at finite temperatures, as evidenced by two finite temperature models: projected stochastic GPE and equilibrium ZNG model. Notably, at lower temperatures and under minimal damping conditions, these oscillations gradually decay until the vortex settles at the centre of the system. In the presence of significant damping, the oscillations are damped.

Supporting these observations, an analytical model of the vortex dynamics was developed in \cite{PhysRevResearch.4.043171}. This model assumes the vortex moves through the rings' central, low-density area and qualitatively describes the damping level at which oscillations vanish. These numerical and analytical findings indicate that the oscillation frequency might serve as a reliable indicator of inherent system properties, and the duration of oscillation could act as a temperature indicator. Given the current state of experimental techniques and detection methods, these results open the way to novel quantum devices and sensors. While a high-damping scenario typically positions the vortex at the system's centre, any external rotation or acceleration would alter the ultimate location of the vortex. This allows potential application of the linked atomic rings in developing an accelerometer, which is a promising direction for future research.

\section{Three-dimensional dynamics of quantum vortices in a coupled systems with curved geometry}
\begin{figure}[h]
\centering
\includegraphics[width=\textwidth   ]{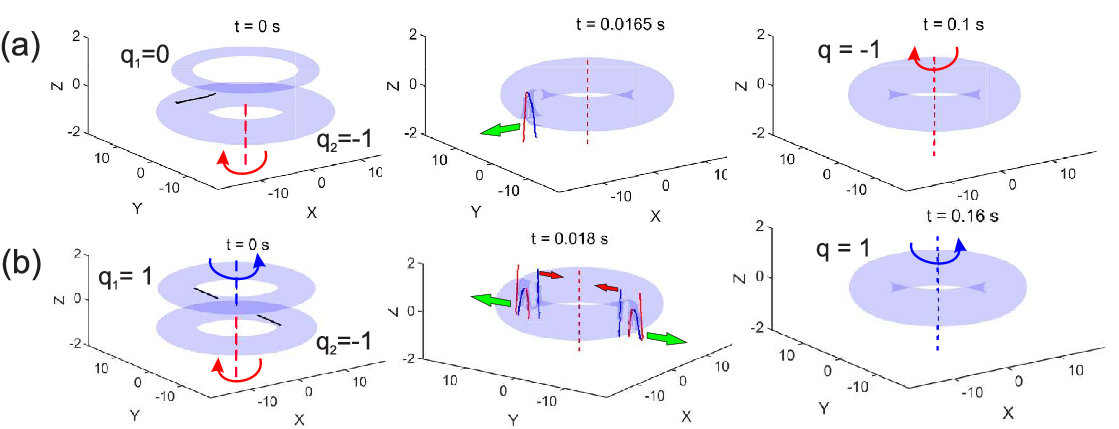}
\caption{ The evolution of the merging squeezed in vertical direction rings as described in Refs. \cite{2019Symm...11.1312O, OLIINYK2020105113}.
 (a) The dynamics of ring merging when the lower ring (characterized by $q_2=-1$) has a significantly larger population than the upper ring ($q_1=0$), denoted as $N_2 \gg N_1$. A key feature of this process is the transformation of the rotational Josephson vortex, initially horizontal (solid black line), into a vertical orientation, eventually forming a V-shaped configuration of a vortex and an antivortex (indicated in blue and red, respectively). This figure also highlights the trajectory of these vortices as they drift towards the outer edge of the system, guided by the direction of the green arrows, resulting in a final state that supports a clockwise persistent current with a quantum number $q=-1$.  (b) Merging of counter-propagating rings with topological charges  $q_1=1$ and $q_2=-1$, where the number of atoms in each ring is slightly different ($N_1 < N_2$). \textcolor{black}{Notably, the symmetric movement of quantum vortices towards the outer boundaries (shown by green arrows), and the pair of vortex lines moving to the inner region (shown by red arrows) dictates that the final state of the system, with a quantum number $q=1$, is determined by the ring with fewer atoms.}}
\label{fig:Mergers2}
\end{figure}

As highlighted in Sec. II, the dynamics of vortices in a flattened (oblate) toroidal BECs can be effectively described using a 2D model. However, the behaviour of vortices undergoes a significant transformation in vertically stacked toroidal condensates, where the vortices exhibit 3D dynamics. In this section, we consider the dynamics of merging persistent currents in coupled BEC rings, with a focus on the influence of three-dimensional nonlinear dynamics of Josephson Vortices (JVs). We discuss how these states can be experimentally created allowing the controllable generation of the vortices in quasi-2D shell-shaped condensates with cylindrical geometry \cite{2017PhRvA..96f3608G}.
\subsection{Rotational Josephson vortices in stacked ring-shaped condensates}\label{subsec:Merging}
The emergence of rotational fluxons, or Josephson vortices, in the low-density regions between coupled toroidal BECs has been studied in Refs. \cite{Oliinyk_2019, OLIINYK2020105113, 2019Symm...11.1312O}. The formation of the rotational Josephson vortices is related to the spontaneous breaking of axial symmetry due to tunnelling flows, which signifies an important property of quantum vortex dynamics in BECs. The formation of these vortices, even within symmetric trapping potentials, effectively breaks rotational symmetry in such quantum systems.

The detailed analysis of the tunnelling phenomena between weakly coupled toroidal BECs demonstrated the impact of population imbalances and barrier modulations on the dynamics and final angular momenta of merging BEC rings.
The behavior of these fluxons leads to varied outcomes for the final state of the condensate. The final state depends on two main factors: the initial population difference between the two rings and the aspect ratio of the toroidal trap in which they are contained.
Unlike classical fluid dynamics, where viscosity dominates, the quantum system exhibits weak dissipation, allowing for the emergence of states with increased angular momentum. Specifically, in elongated condensates, the vorticity of the final state aligns with that of the more populated ring. However, in "pancake-shaped" rings, there exists a critical imbalance level; below this level, the system evolves to a state with no net vorticity, whereas above it, the final state inherits the vorticity of the dominant ring component. Two examples of evolution for merging "pancake-shaped" rings are shown in Fig. \ref{fig:Mergers2}. \textcolor{black}{The details of the relaxation process and the role of symmetry in such systems were previously investigated in our work \cite{2019Symm...11.1312O}.}

Employing the damped GPE (\ref{GPE}), the dynamics of BECs and the eventual stabilization of Josephson vortices towards equilibrium states within toroidal BECs was investigated. The toroidal potential (\ref{toroidal_trap}) was combined with a repulsive barrier $V_b=U(t)\exp\left(-\frac12(z-z_0)^2/a^2\right)$, describing a blue-detuned sheet beam, which splits the toroidal condensate in upper and lower weakly coupled rings-shaped components. Note that, by shifting centre $z_0$ of the splitting barrier, it is easy to prepare an initial state with a dominant population in the ring with topological charge $q_1$ ($N_1 > N_2$ for $z_0 > 0$) or $q_2$ ($N_2 > N_1$ for $z_0 < 0$).

Further investigations \cite{Oliinyk_2019,OLIINYK2020105113,2019Symm...11.1312O} have demonstrated how variations in barrier properties and initial conditions can lead to diverse outcomes, including distinct patterns of symmetry breaking, vortex formation, and superfluid current distributions.

\subsection{Methods of generation of Josephson vortices in stacked toroidal condensates}\label{subsec:create_stacked}
\begin{figure}[htb]
\centering
\includegraphics[width=\textwidth   ]{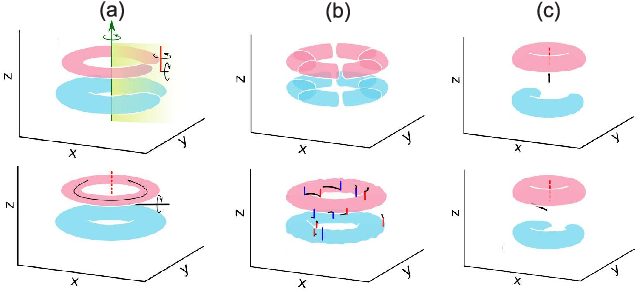}
\caption{Three methods to prepare states with different topological charges in two coupled coaxial ring-shaped atomic Bose-Einstein condensates as described in Ref. \cite{PhysRevA.106.043305}. (a) Generating persistent currents by stirring a system of coaxially stacked toroidal condensates, divided by a splitting potential, with the initial $(0, 0)$ state shown in the upper row. The rings contain different numbers of atoms, and the stirring potential, depicted by the yellow area, rotates anticlockwise with constant angular velocity. \textcolor{black}{ (b) Stochastic generation of a persistent currents in weakly coupled rings, resulting from the merging of initially separated fragments with random phase differences. Rapid switching off of the vertical barriers leads to the formation of the double-ring system with a large number of vortices and antivortices. (c) Asymmetric persistent current decay in a double-ring system, where a vortex leaves the lower ring with the simultaneous formation of a Josephson vortex between the rings (black line).}}
\label{fig:Create_stacked_JV}
\end{figure}

The remarkable finding of the work \cite{PhysRevA.106.043305} is the theoretical demonstration of deterministic discontinuous transitions between distinct circulation states in toroidally confined Bose-Einstein condensates. These phase slips are triggered by vortex excitations due to a rotating weak link, leading to the formation of Josephson Vortices (JVs) between the coupled condensates.
Using numerical simulations of 3D damped GPE model, three experimentally accessible methods to generate and manipulate these states within two coupled coaxial ring-shaped atomic BECs were demonstrated in Ref. \cite{PhysRevA.106.043305}.  Each method which reveals different aspects of superflow interactions and JV dynamics are illustrated in Fig. \ref{fig:Create_stacked_JV}.

Stirring of the asymmetrically populated double-ring system: This method employs a stirring laser beam to generate persistent currents in one of the rings, particularly effective in systems with population imbalances. It is based on the density differences between the rings to selectively induce superflows (see Fig. \ref{fig:Create_stacked_JV} (a)).

Stochastic generation of Josephson vortices: This method involves merging initially separated condensate fragments with random phase differences, leading to the spontaneous formation of JVs, as illustrated in Fig. \ref{fig:Create_stacked_JV} (b). This stochastic approach provides insights into the role of phase coherence and the emergent behavior of superfluid systems under conditions of controlled disorder.

Asymmetric decay of persistent current in a double-ring system is shown in \ref{fig:Create_stacked_JV} (c): This approach involves initially persistent currents with the same topological charges in both rings and then selectively dissipating the current in one ring via a controlled perturbation. This method highlights the role of external potentials in manipulating the decay pathways of persistent currents.

Each of these methods opens a way for creating a versatile platform for investigating phenomena ranging from Josephson effects in weakly interacting regimes to quantum Kelvin-Helmholtz instability in merging rings. Moreover, the study illuminates the potential for using rotational JVs in atomic Josephson junctions as prospective tools for precision rotation measurements with atomic matter waves.

\subsection{Stable 3D vortex structures in curved geometries}\label{subsec:create_hybrid}
\begin{figure}[htb]
\centering
\includegraphics[width=\textwidth   ]{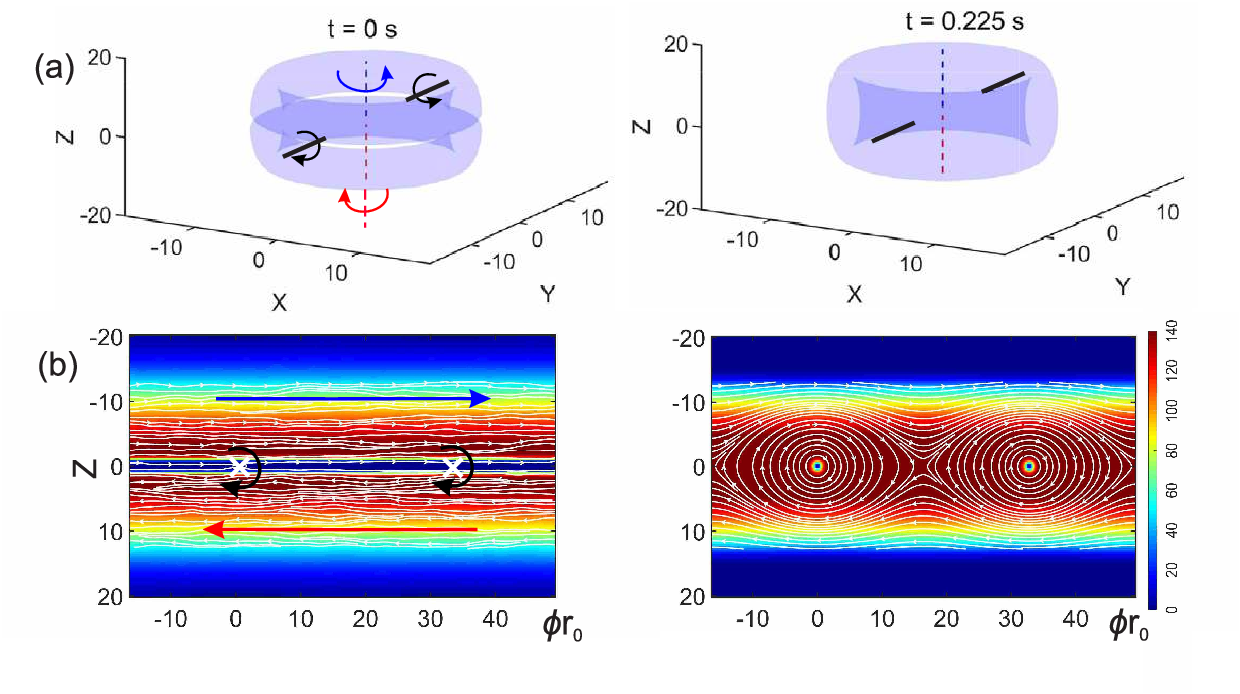}
\caption{(a) Snapshots at two moments, illustrating the evolution of the merging strongly elongated (prolate) toroidal condensates as described in Ref. \cite{OLIINYK2020105113}. Shown are 3D isosurfaces with constant condensate density, and (b) maps of the distribution of the density and condensate flows on the cylindrical surface of radius $r_0$, with $\phi$ representing the angular coordinate. A long-lived hybrid complex with cylindrical geometry emerges from the evolution of the condensate under the influence of weak dissipation. Blue (red) arrows show the directions of vortex (antivortex) flows in the upper (lower) rings. Horizontal black lines represent Josephson vortex cores.
}
\label{fig:Mergers}
\end{figure}
In the study of interacting BECs within vertically stacked toroidal traps, a notable observation is the formation of long-lived hybrid vortex structures exhibiting hidden vorticity \cite{OLIINYK2020105113}. These configurations arise in systems of two coaxial BEC rings that are axially separated by a potential barrier, each ring carrying opposite vorticities but collectively maintaining a composite zero angular momentum.

The emergence of  $|q_1-q_2|$ fluxons at the interface between superfluid flows with different topological charges, $q_1$ and $q_2$, is an inevitable consequence of the azimuthal periodicity intrinsic to the toroidal condensate wave function. The existence of JVs guarantees that the wave functions of each ring with different topological charges are single-valued.

It turns out that dynamics of the quantum vortices, observed after the merger of the rings, crucially depend on the aspect ratio $A=\omega_z/\omega_r$. For BECs of an elongated trap $A<1$, the energetically favourable configuration corresponds to JVs aligning horizontally, even in the absence of a separating barrier between the rings. Using simple energetic estimates and direct numerical simulation of dissipative 3D GPE, it was demonstrated in Ref.  \cite{OLIINYK2020105113}  that rotational fluxons are most likely to keep the horizontal orientation when elongated rings merge. Furthermore, this property of elongated condensates does not depend on the number of atoms but is determined by the aspect ratio of the trap only.

Figure \ref{fig:Mergers} illustrates the density and flows of the condensate corresponding to the combination of four vortex lines. Vortex lines with a vertical orientation are trapped within the potential well formed by the toroidal condensate. The hybrid vortex structures are characterized by their topological properties, including vertically aligned vortex lines with horizontally oriented JVs.
It is remarkable that instead of the development of the classical Kelvin–Helmholtz
instability at the interface of the merging persistent currents in a prolate
potential trap, sufficiently elongated in the axial direction ($A<1$), one observes the formation of nonlinear robust hybrid vortex structures (as
illustrated in Fig. \ref{fig:Mergers}). The dynamics of these hybrid structures are investigated in Ref. \cite{OLIINYK2020105113} under various conditions, including population imbalances between the rings and different interaction regimes ranging from weak (characterized by tunnelling through the barrier) to strong (where the rings merge across a reduced barrier).

\section{Conclusions and prospects}
In conclusion, we have synthesized recent works on quantum vortices in atomic BECs subject to curved geometric constraints.
Comparing various BEC systems we elucidate topological and dimensional differences, which lead to unique physical manifestations with remarkable implications for the dynamics and stability of vortex structures in curved geometries. The toroidal topology of the trapped BEC, which we have considered, guarantees that vortices are pinned at the centre of the ring-shaped condensate, where the vortex energy has a local minimum, so that vortex states can be very robust. Furthermore, fundamental properties, dynamics and stability of topological excitations in BEC can be significantly affected by the curved geometry in low-dimensional systems.  A dark soliton is inherent to the quasi-1D system, and it is characterized by a node of the condensate density, accompanied by a phase jump across the soliton. The emergence of the dark solitons was illustrated by simulations of quasi-1D BEC in elliptic waveguide \cite{2023NJPh...25j3003N}.  In higher dimensions, such as in a 2D system, a vortex carries quantized angular momentum and is associated with a phase winding around its core. In a purely 2D system, the vortex core is a point, while in 3D systems, the vortex core may either be a line that starts and ends at the condensate's boundary or a closed-loop ring. The energy of the vortex increases with the length of the vortex core. Thus in a pancake-shape condensate, a short vortex line keeps the vertical orientations and its behaviour can be effectively described in the framework of the 2D model. Experimental evidence of this property of the vortices in an oblate BEC confined in an annular trapping potential is inverse energy cascade which is inherent for 2D quantum turbulence \cite{PhysRevLett.111.235301}. However, if a long vortex line appears in the bulk of the condensate it starts bending and drifts to the lower-density regions minimizing its energy. {For instance, in an oblate system, the vortex line can rapidly change its orientation as illustrated in Fig. \ref{fig:Mergers2} for rotational Josephson vortices in merging ring-shaped condensates.  } These findings open the way for the generation of the vortices in very prolate quasi-2D shell-shaped condensates (with almost cylindrical geometry) allowing experimental demonstration of the single and multiple vortex structures revealed in Ref. \cite{2017PhRvA..96f3608G}. Furthermore, as was demonstrated in Ref. \cite{OLIINYK2020105113} using a pinning potential it is possible to create a completely stable stationary 3D hybrid vortex complexes with a tunable value of the angular momentum per particle in the range of $q_1 < L_z/N < q_2$. Such stable hybrid structures with controllable vorticity in curved geometry open a way for potential applications in atomtronic circuits and quantum information processing such as topologically protected qubits \cite{PhysRevLett.94.166802}, where control over quantum flow and coherence is essential.

The prospects for further studies in this field are vast, promising exciting developments in the realm of quantum technologies, as we continue to probe the frontiers of quantum physics in curved spatial configurations. Future studies on vortices in BECs in curved geometries could focus on the formation, stability and dynamics of quantized vortices in two-dimensional curved superfluids. These studies could deepen our understanding of superfluid behaviour in non-Euclidean geometries and reveal remarkable properties of vortex dynamics influenced by curvature.  In general, the increasing technical capability of fabricating condensed matter systems with nanoscale-level control \cite{Gentile2022} allows for investigations of the geometry, and curvature impacts on their quantum many-body physics. In perspective, this progress enables both fundamental investigations of quantum physics in curved spatial domains and technological applications, in which the curved geometry allows engineering and control of the properties of materials \cite{2010RvMP...82.1301T,
2002dgcm.book.....N,Streubel_2016,Peng2020StrainEngineering, Gentile2022}.

Reaching the two-dimensional shell regime in the experiments of Ref. \cite{PhysRevLett.129.243402} could help to investigate vorticity and superfluid hydrodynamics in a fully-closed spherical surface. In addition, the fact that this geometry is obtained with phase-separated repulsive mixtures could allow for studies of vortices with massive cores in the ellipsoidal geometry \cite{2020PhRvA.101a3630R}.
Also, performing the experiments in ground-based laboratories may provide practical advantages over microgravity experiments and simplify both the production of shells, their imaging and the collection of large amounts of data.
At the same time, the microgravity experimental facilities on the International Space Station will soon allow for a larger number of atoms and better control, enabling the production of fully closed 2D condensate shells. This can open the way to the exploration of vortex dynamics in the ellipsoidal geometry \cite{PhysRevA.105.023307}.

The introduction of external fields and interactions, such as spin-orbit coupling in curved geometries, could lead to exotic vortex states with non-trivial topology \cite{PhysRevA.84.063604,PhysRevA.109.013326}. The study of vortex matter on topologically non-trivial surfaces, like M\"obius strips, might uncover unusual superfluid phases and symmetry-breaking patterns \cite{PhysRevLett.129.016801,2022PRXQ....3a0316L}. Studies of the quantum Hall effect on curved surfaces could bridge the gap between flat and curved quantum systems, providing a new understanding of quantum coherence in curved spaces \cite{PhysRevA.96.033622}. Moreover, the development of advanced numerical simulations and topological field theories tailored for curved geometries could significantly enhance our theoretical understanding and predictive capabilities. These studies, stimulating emerging experimental techniques to create and manipulate quantum gases on designed curved surfaces, would not only test these theoretical predictions but also open new ways to novel quantum technologies based on the unique properties of curved-space quantum systems \cite{2021AVSQS...3c9201A}.

Studies of the vortex excitations in curved geometries have opened new prospects in various physical systems, well beyond atomic BECs.
Among the myriad hypotheses about the elusive nature, composition, and physical characteristics of dark matter (DM), the proposition that DM is composed of ultralight bosons, which assume the state of BECs, emerges as a compelling solution \cite{2021A&ARv..29....7F}.
Previous investigations have demonstrated the existence and stability of the vortex structures in the BEC solitonic core, which can exist in the central regions of the galactic halos (see, e.g. \cite{PhysRevD.108.023503,2021LTP....47..684N,2023EPJC...83..451K}). Future studies of the vortices in the ultralight bosonic dark matter can involve analysis of the space-time curvature and frame-dragging effects in the vicinity of rotating super-massive black holes. Analogue physics offers a unique platform for emulating inaccessible phenomena specific to curved space-time in the vicinity of black holes, enabling their study within the controlled environment of laboratory experiments \cite{vsvanvcara2024rotating}. In perspective, analog simulations of cosmological phenomena, such as the inflationary expansion of the early Universe, using expanding superfluid shells, present a fascinating avenue for connecting quantum fluid dynamics with cosmology \cite{PhysRevX.8.021021}.

\textcolor{black}{We have studied the primary factors influencing vortex dynamics in BECs within curved geometries, including dimensionality, topology, non-constant curvature and density inhomogeneity. Dimensionality allows the formation of different types of excitations in curved waveguides, with notable variations between quasi one-dimensional and three-dimensional systems, as was pointed out in Sec. II A. Topologically, the contrast between toroidal and single-connected configurations introduces distinct constraints that impact vortex stability and behaviour, as was highlighted in Sec. II B and C. The presence of non-constant curvature, discussed in Sec. II A, adds complexity by creating spatially varying potential landscapes. Furthermore, density inhomogeneities, such as those induced by weak links, result in local variations that significantly affect vortex dynamics, as was illustrated in Sec. II B. It is important, that these effects can compete and interact. For example, as was pointed out in Sec. II A, the toroidal topology combined with curvature-induced gradient of the phase can significantly modify dynamics and stability of the persistent current in the ring-shaped condensate. Another example: density modulation, which appear as the result of the curvature, can be smoothed out by variation of the potential profile, which smooth-out the density along the waveguide, as was experimentally demonstrated in \cite{Campo2014}. Therefore, competition and interplay between dimensionality, topology, non-constant curvature and density inhomogeneity drive the intricate nature of vortex behaviour in curved BECs, highlighting perspectives for future research and applications in quantum technologies.
}

\begin{acknowledgments}
A. T. acknowledges support from ANR grant ``Droplets'' No.~ANR-19-CE30-0003-02 and from the EU Quantum Flagship (PASQuanS2.1, 101113690).
A.Y. acknowledges support from BIRD Project “Ultracold atoms in curved geometries” of the University
of Padova.
LS is partially supported by the European Union-NextGenerationEU within
the National Center for HPC, Big Data and Quantum Computing
[Project No. CN00000013, CN1 Spoke 10: Quantum Computing] and by the European Quantum Flagship Project PASQuanS 2. L.S. and A.Y. are also partially supported by Iniziativa Specifica Quantum of Istituto Nazionale di Fisica Nucleare, by the Project Frontiere Quantistiche within the 2023 funding programme  'Dipartimenti di Eccellenza' of the Italian Ministry for Universities and Research, and by the PRIN 2022 Project Quantum Atomic Mixtures: Droplets, Topological Structures, and Vortices.

ICFO group acknowledges support from:
Europea Research Council AdG NOQIA;
MCIN/AEI  (PGC2018-0910.13039 /501100011033, EX2019-000910-S/10.13039/501100011033, Plan National FIDEUA PID2019-106901GB-I00, Plan National STAMEENA PID2022-139099NB, I00, project funded by MCIN/AEI/10.13039/501100011033 and by the “European Union NextGenerationEU/PRTR" (PRTR-C17.I1), FPI); QUANTERA MAQS PCI2019-111828-2); QUANTERA DYNAMITE PCI2022-132919, QuantERA II Programme co-funded by European Union’s Horizon 2020 program under Grant Agreement No 101017733);
Ministry for Digital Transformation and of Civil Service of the Spanish Government through the QUANTUM ENIA project call - Quantum Spain project, and by the European Union through the Recovery, Transformation and Resilience Plan - NextGenerationEU within the framework of the Digital Spain 2026 Agenda;
Fundació Cellex;
Fundació Mir-Puig;
Generalitat de Catalunya (European Social Fund FEDER and CERCA program, AGAUR Grant No. 2021 SGR 01452, QuantumCAT \ U16-011424, co-funded by ERDF Operational Program of Catalonia 2014-2020);
Barcelona Supercomputing Center MareNostrum (FI-2023-1-0013);
Funded by the European Union. Views and opinions expressed are however those of the author(s) only and do not necessarily reflect those of the European Union, European Commission, European Climate, Infrastructure and Environment Executive Agency (CINEA), or any other granting authority.  Neither the European Union nor any granting authority can be held responsible for them (EU Quantum Flagship PASQuanS2.1, 101113690, EU Horizon 2020 FET-OPEN OPTOlogic, Grant No 899794),  EU Horizon Europe Program (This project has received funding from the European Union’s Horizon Europe research and innovation program under grant agreement No 101080086 NeQSTGrant Agreement 101080086 — NeQST);
ICFO Internal “QuantumGaudi” project;
European Union’s Horizon 2020 program under the Marie Sklodowska-Curie grant agreement No 847648;
“La Caixa” Junior Leaders fellowships, La Caixa” Foundation (ID 100010434): CF/BQ/PR23/11980043.

\end{acknowledgments}

\section*{DATA AVAILABILITY STATEMENT}
The data that support the findings of this study are available from the corresponding author upon reasonable request.

\section*{Author Declarations}
\subsection*{Conflict of interest}
The authors have no conflicts to disclose.

\bibliographystyle{apsrev4-2}
\bibliography{Refs}

\end{document}